\documentclass{emulateapj}

\newcommand{\farcsec}{\mbox{\ensuremath{.\!\!^{\mbox{\scriptsize{s}}}}}}

\newcommand{\degree}{\ensuremath{^\circ}}

\slugcomment{Accepted for publication in The Astrophysical Journal}

\shorttitle{PROPER MOTION FOR PWN~G359.23$-$0.82}
\shortauthors{Hales et al.}

\begin{document}


\title{A Proper Motion for the Pulsar Wind Nebula G359.23$-$0.82, ``the
Mouse,'' Associated with the Energetic Radio Pulsar J1747$-$2958}


\author{C. A. Hales\altaffilmark{1,2}, B. M. Gaensler\altaffilmark{1,3}, S. Chatterjee\altaffilmark{1,4},
E. van der Swaluw\altaffilmark{5}, and F. Camilo\altaffilmark{6}}
\altaffiltext{1}{Sydney Institute for Astronomy (SIfA), School of Physics, The University of Sydney, NSW 2006, Australia;
c.hales@physics.usyd.edu.au}
\altaffiltext{2}{CSIRO Australia Telescope National Facility, Epping NSW 1710, Australia}
\altaffiltext{3}{Australian Research Council Federation Fellow}
\altaffiltext{4}{Current address: Department of Astronomy, Cornell University, Ithaca, NY 14853}
\altaffiltext{5}{National Institute for Public Health and the Environment (RIVM), 3720 BA Bilthoven, The Netherlands}
\altaffiltext{6}{Columbia Astrophysics Laboratory, Columbia University, New York, NY 10027}




\begin{abstract}

The ``Mouse'' (PWN~G359.23$-$0.82) is a spectacular bow shock pulsar wind nebula, powered by the radio pulsar J1747$-$2958. The
pulsar and its nebula are presumed to have a high space velocity, but their proper motions have not been directly measured.
Here we present 8.5 GHz interferometric observations of the Mouse nebula with the Very Large Array, spanning a time
baseline of 12 yr. We measure eastward proper motion for PWN~G359.23$-$0.82 (and hence indirectly for PSR~J1747$-$2958)
of $12.9\pm1.8$~mas~yr$^{-1}$, which at an assumed distance of 5~kpc corresponds to a transverse space velocity
of $306\pm43$~km~s$^{-1}$. Considering pressure balance at the apex of the bow shock,
we calculate an in situ hydrogen number density of approximately $1.0_{-0.2}^{+0.4}$~cm$^{-3}$ for the interstellar medium
through which the system is traveling. A lower age limit for PSR~J1747$-$2958 of $163_{-20}^{+28}$~kyr is calculated by considering
its potential birth site. The large discrepancy with the pulsar's spin-down age of 25~kyr is possibly explained by surface dipole
magnetic field growth on a timescale $\approx$15~kyr, suggesting possible future evolution of PSR~J1747$-$2958 to a different class
of neutron star. We also argue that the adjacent supernova remnant G359.1$-$0.5 is not physically associated
with the Mouse system but is rather an unrelated object along the line of sight.

\end{abstract}


\section{Introduction}\label{SectionIntroduction}

The evolution of neutron stars and the potential relationships between some of their observed classes remain outstanding problems in
astrophysics. Proper motion studies of neutron stars can provide independent age estimates with which to shed light on these questions.
In particular, the well defined geometry of bow shock pulsar wind nebulae (PWNe; \citeauthor{gaensler:3} \citeyear{gaensler:3}),
where the relativistic wind from a high-velocity pulsar is confined by ram pressure, can be used as a probe to aid in the understanding
of both neutron star evolution and the properties of the local medium through which these stars travel.

\begin{figure*}[t]
\setlength{\abovecaptionskip}{-7pt}
\begin{center}
\includegraphics[trim = 0mm 0mm 0mm 0mm, clip, angle=-90, width=13cm]{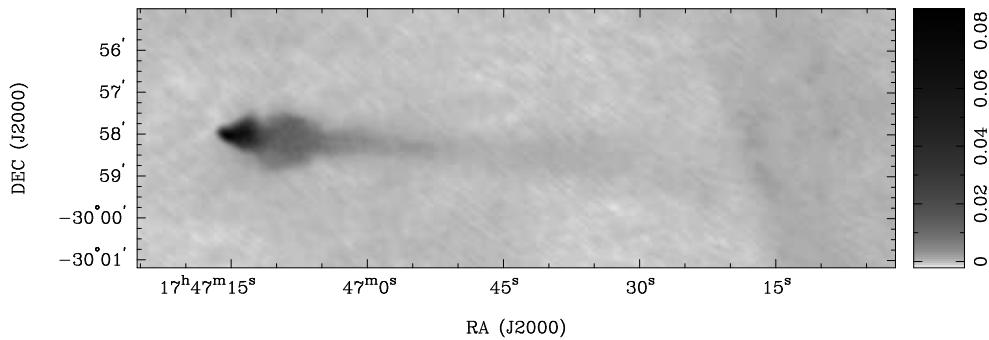}
\end{center}
\caption{VLA image of the Mouse (PWN~G359.23$-$0.82) at 1.4~GHz with a resolution of 12\farcs8$\times$8\farcs4 (reproduced from
\citeauthor{gaensler:5} \citeyear{gaensler:5}). The brightness scale is logarithmic, ranging between $-$2.0
and $+$87.6~mJy~beam$^{-1}$ as indicated by the scale bar to the right of the image. The eastern rim of SNR~G359.1$-$0.5 is faintly visible west
of $\sim$RA~17$^{\mbox{\scriptsize{h}}}$46$^{\mbox{\scriptsize{m}}}$25$^{\mbox{\scriptsize{s}}}$.}
\label{fig:yusef20cm}
\end{figure*}

The ``Mouse'' (PWN~G359.23$-$0.82), a non-thermal radio nebula, was discovered as part of a radio continuum survey of the
Galactic center region \citep{yusef}, and was suggested to be powered by a young pulsar following X-ray detection \citep{predehl}.
It is now recognized as a bow shock PWN moving supersonically through the interstellar medium (ISM; \citeauthor{gaensler:5} \citeyear{gaensler:5}).
Its axially symmetric morphology, shown in Figure \ref{fig:yusef20cm}, consists of a compact ``head'', a fainter ``body'' extending
for $\sim$10$^\prime$$^\prime$, and a long ``tail'' that extends westward behind the Mouse for $\sim$40$^\prime$$^\prime$
and $\sim$12$^\prime$ at X-ray and radio wavelengths respectively \citep{gaensler:5,mori}. The cometary tail appears to indicate
motion away from a nearby supernova remnant (SNR), G359.1$-$0.5 \citep{yusef}.

A radio pulsar, J1747$-$2958, has been discovered within the ``head'' of the Mouse \citep{camilo:1}. PSR~J1747$-$2958 has a spin
period $P=98.8$ ms and period derivative $\dot{P}=6.1\times10^{-14}$, implying a spin-down luminosity
$\dot{E}=2.5 \times 10^{36}$~ergs~s$^{-1}$, surface dipole magnetic field strength $B=2.5\times10^{12}$~G, and characteristic age
$\tau_{c} \equiv P/ 2\dot{P}=25$~kyr (\citeauthor{camilo:1} \citeyear{camilo:1}; see also updated timing data from
\citeauthor{gaensler:5} \citeyear{gaensler:5}). The distance to the pulsar is {\footnotesize $\gtrsim$}4~kpc from X-ray
absorption \citep{gaensler:5}, and {\footnotesize $\lesssim$}5.5 kpc~from HI absorption \citep{uchida}. Here we assume that
the system lies at a distance of $d=5d_{5}$~kpc, where $d_{5}=1\pm0.2$ ($1\sigma$).

Given such a small characteristic age, it is natural to ask where PSR~J1747$-$2958 was born and to try and find an
associated SNR. While it is possible that no shell-type SNR is visible, such as with the Crab pulsar \citep{sankrit} and other young
pulsars \citep{braun}, an association with the adjacent SNR~G359.1$-$0.5 appears plausible. This remnant was initially suggested to
be an unrelated background object near the Galactic center \citep{uchida}. However, it is now believed that the two may be located at
roughly the same distance (\citeauthor{yusef3} \citeyear{yusef3}, and references therein). By determining a proper motion for
PSR~J1747$-$2958, this association can be subjected to further scrutiny (for example, see analysis of PSR~B1757$-$24,
PWN~G5.27$-$0.90 and SNR~G5.4$-$1.2; \citeauthor{blazek} \citeyear{blazek}; \citeauthor{zeiger} \citeyear{zeiger}).

As PSR~J1747$-$2958 is a very faint radio source, it is difficult to measure its proper motion interferometrically. It is also
difficult to use pulsar timing to measure its proper motion due to
timing noise and its location near the ecliptic plane \citep{camilo:1}. To circumvent these issues, in this paper we investigate
dual-epoch high-resolution radio observations of the Mouse nebula, spanning 12 years from 1993
to 2005, with the intention of indirectly inferring the motion of PSR~J1747$-$2958 through the motion of its bow shock PWN. In
\S~\ref{SectionObservations} we present these observations. In \S~\ref{SectionAnalysis} we present our analysis and measurement
of proper motion using derivative images of PWN~G359.23$-$0.82. In \S~\ref{SectionDiscussion} we use our measurement to determine an
in situ hydrogen number density for the local ISM, to resolve the question of a possible association with SNR~G359.1$-$0.5, and to investigate
the age and possible future evolution of PSR~J1747$-$2958. We summarize our conclusions in \S~\ref{SectionConclusions}.

\section{Observations}\label{SectionObservations}

PWN~G359.23$-$0.82 was observed with the Very Large Array (VLA) on 1993 February 2 (Program AF245) and again\footnote{An observation of
PWN~G359.23$-$0.82 was also carried out on 1999 October 8 (Program AG571). However, the target was observed mainly at low elevation and
the point spread function and spatial frequency coverage were both poor as a result, thus ruling out the observation's amenability to
astrometric comparison.} on 2005 January 22 (Program AG671). Each of these observations were carried out in the hybrid BnA configuration
at a frequency near 8.5 GHz. The 1993 and 2005 epochs used on-source observation times of 3.12 and 2.72 hours respectively. The 1993
observation only measured $RR$ and $LL$ circular polarization products, while the 2005 observation measured the cross terms $RL$ and
$LR$ as well. Both observations used the same pointing center, located at
$\mbox{RA}$~$=$~$17^{\mbox{\scriptsize{h}}}$47$^{\mbox{\scriptsize{m}}}$15\farcsec764, $\mbox{Dec}$~$=$~$-29$\degree58$^\prime$1\farcs12 (J2000),
as well as the same primary flux calibrator, 3C286. Both were phase-referenced to the extragalactic source TXS~1741$-$312, located at
$\mbox{RA}$~$=$~$17^{\mbox{\scriptsize{h}}}$44$^{\mbox{\scriptsize{m}}}$23\farcsec59, $\mbox{Dec}$~$=$~$-31$\degree16$^\prime$35\farcs97 (J2000),
which is separated by $1\hbox{$.\!\!^\circ$}4$ from the pointing center.

\begin{figure*}[t]
\setlength{\abovecaptionskip}{-7pt}
\begin{center}
\includegraphics[trim = 0mm 0mm 0mm 0mm, clip, angle=-90, width=12cm]{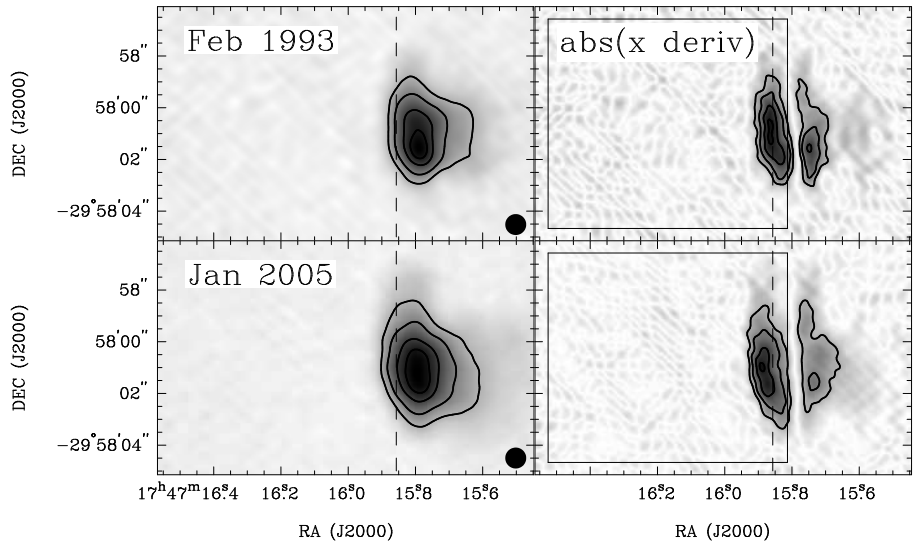}
\end{center}
\caption{{\it Left column:} VLA observations of the Mouse at 8.5~GHz over two epochs separated by 12 yr. Each image covers
a 14\farcs5$\times$8\farcs5 field at a resolution of 0\farcs81, as indicated by the circle at the bottom right of each panel. The brightness
scale is linear, ranging between $-$0.23 and $+$3.5~mJy~beam$^{-1}$. 
{\it Right column:} Spatial x-derivative images for 1993 (top) and 2005 (bottom), covering the same regions as images in the left column.
The images are shown in absolute value to increase visual contrast using a linear brightness scale spanning zero and the largest magnitude
derivative from either image. The box between $\sim$RA~17$^{\mbox{\scriptsize{h}}}$47$^{\mbox{\scriptsize{m}}}$15\farcsec8$-$16\farcsec5 indicates
the eastern region extracted for cross-correlation.
{\it All panels:} Contours are shown over each image at 30\%, 50\%, 75\% and 90\% of the peak flux within their respective column group.
The dashed vertical line in each panel has been arbitrarily placed at a right ascension near the brightness peak in the top-right panel
in order to determine if motion can be seen by eye between epochs.}
\label{fig:allImages}
\end{figure*}

Data reduction was carried out in near identical fashion for both epochs using the MIRIAD package \citep{sault:2}, taking into
consideration the slightly different correlator mode used in the 1993 data. This process involved editing, calibrating, and
imaging the data using multi-frequency synthesis and square pixels of size 50~$\times$~50 milli-arcseconds. These images were
then deconvolved using a maximum entropy algorithm and smoothed to a common resolution with a circular Gaussian of full width at
half-maximum (FWHM) 0\farcs81. The resulting images are shown in the left column of Figure \ref{fig:allImages}.

The peak flux densities of the 1993 and 2005 images are 3.24 and 3.25~mJy~beam$^{-1}$, respectively; the noise in these two images are
51 and 35~$\mu$Jy~beam$^{-1}$, respectively. The pulsar J1747$-$2958 is located at
$\mbox{RA}$~$=$~$17^{\mbox{\scriptsize{h}}}$47$^{\mbox{\scriptsize{m}}}$15\farcsec882,
$\mbox{Dec}$~$=$~$-29$\degree58$^\prime$1\farcs0 (J2000), within the region of intense synchrotron emission seen
in each image (see \S~3.5 of \citeauthor{gaensler:5} \citeyear{gaensler:5}). Qualitatively comparing each epoch from the left column of
Figure \ref{fig:allImages}, it appears that the head of PWN~G359.23$-$0.82 has the same overall shape in both images,
with a quasi-parabolic eastern face, approximate axial symmetry along a horizontal axis through the centre of the nebula (although
the position of peak intensity seems to shift slightly in declination), and a small extension to the west. By eye the PWN seems
to be moving from west to east over time, in agreement with expectation from the cometary morphology seen in Figure \ref{fig:yusef20cm}.
Beyond any minor morphological changes seen between the images in the left column of Figure \ref{fig:allImages}, the Mouse nebula seems
to have expanded slightly over time.

\section{Analysis}\label{SectionAnalysis}

To quantify any motion between epochs, an algorithm was developed to evaluate the cross-correlation coefficient over a range of
pixel shifts between images, essentially by producing a map of these coefficients. This algorithm made use of the Karma visualization package
\citep{gooch} to impose accurate non-integer pixel shifts.

We applied our algorithm to the image pair from the left column of Figure \ref{fig:allImages} to determine an image offset measurement. To check
that this offset measurement would be robust against any possible nebular morphological change between epochs, we also applied our algorithm to the
same image pair when both images were equally clipped at various flux density upper-level cutoffs. We found that the offset measurement was strongly
dependent on the choice of flux density cutoff. Clearly such variation in the measured shift between epochs was not desirable, as selection of
a final solution would have required an arbitrary assumption about the appropriate level of flux density clipping to use. There was also no
strong indication that the region of peak flux density in each image coupled with the exact location of PSR~J1747$-$1958. In order to isolate
the motion of the pulsar from as much nebular morphological change as possible, we focused on a different method involving the cross-correlation
of spatial derivatives of the images from the left column of Figure \ref{fig:allImages}.

As there is not enough information to solve for an independent Dec-shift, we will only focus on an RA-shift, and will assume that any Dec changes
are primarily due to morphological evolution of the Mouse nebula. To justify this assumption, we note simplistically that the cometary tail
of the Mouse in Figure \ref{fig:yusef20cm} is oriented at {\footnotesize $\lesssim$}5\degree~from the RA-axis, and thus estimate that any
Dec motion contributes less than 10\% to the total proper motion of the nebular system. The small angle also justifies our decision
not to calculate derivatives along orthogonal axes rotated against the RA-Dec coordinate system.

For each epoch, an image of the first spatial derivative of intensity in the x (RA) direction was created by shifting the original
image by 0.1 pixels along the RA-axis, subtracting the original from this shifted version, and then dividing by the value of the shift.
These x-derivative images are shown in the right column of Figure \ref{fig:allImages}, where the brighter pixels represent regions of larger
x-derivative from the corresponding left column images (note that these derivative images are shown in absolute value so as to increase their
visual contrast; this operation was not applied to the analyzed data).

The x-derivative images have signal-to-noise ratios $\sim$13, since progression to higher derivatives degrades
sensitivity. As seen in the right column of Figure \ref{fig:allImages}, the x-derivative images of the Mouse are
divided into two isolated regions: an eastern forward region and a western rear region. Derivatives in the eastern region
are greater in magnitude than those in the western region.

As we will justify in \S~\ref{SectionDiscussion}, we propose that the eastern region of the x-derivative images tracks the forward
termination shock of the Mouse nebula, which in turn acts as a proxy for an upper limit on the motion of PSR~J1747$-$2958. The
eastern region provides a natural localized feature at each epoch with which to generate cross-correlation maps in order to track the motion
of PSR~J1747$-$2958.

\subsection{Calculation of Nebular Proper Motion}\label{Subsection:FinalCalc}

\begin{figure*}[t]
\setlength{\abovecaptionskip}{-3pt}
\begin{center}
\begin{minipage}[c]{0.45\textwidth}
\begin{center}
\vspace{-4mm}
\includegraphics[trim = 0mm 0mm 5mm 0mm, width=5.3cm, angle=-90]{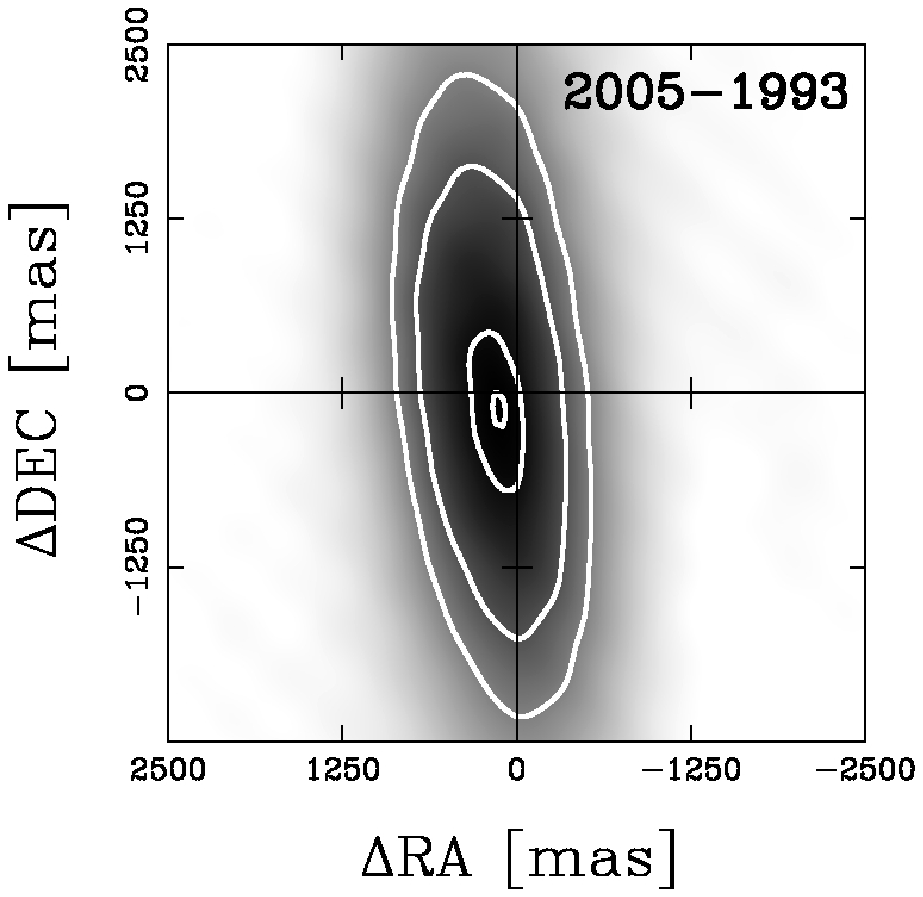}
\end{center}
\end{minipage}
\hspace{1mm}
\begin{minipage}[c]{0.45\textwidth}
\begin{center}
\vspace{-7mm}
\includegraphics[trim = -3mm 0mm 6mm 0mm, width=4.7cm, angle=-90]{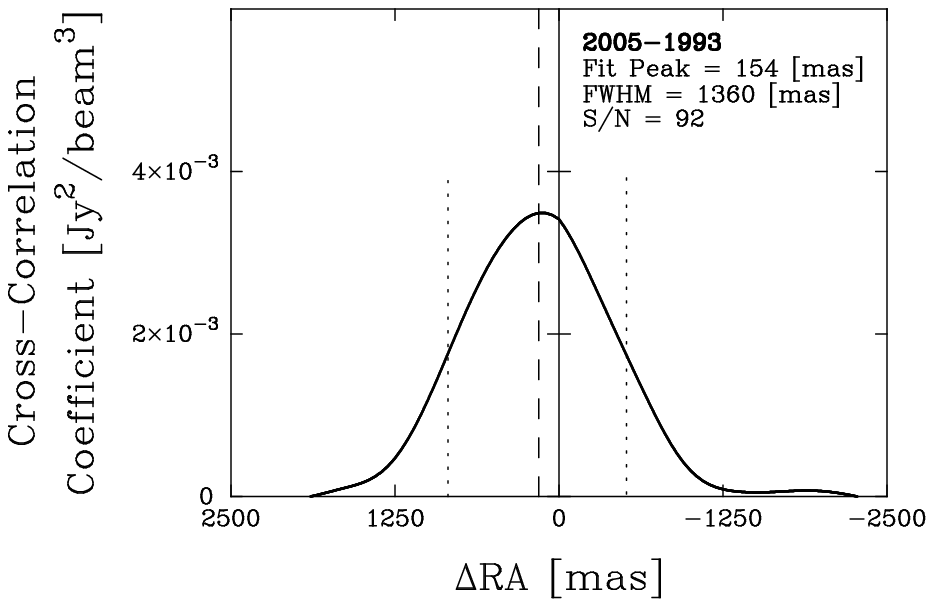}
\end{center}
\end{minipage}
\end{center}
\caption{{\it Left}: Cross-correlation map for cropped and padded x-derivative images between 1993 and 2005. Shifts range
from $-$2500 to $+$2500~mas in both RA and Dec. The pixels are scaled from zero (white) to the peak value of the
map (black), and contours are at 50\%, 68.3\%, 99.5\%, and 99.7\% of this peak value. {\it Right}: Profile along a line
parallel to the RA-axis through the peak value of the cross-correlation map (solid curve). The caption quantifies
the RA-shift for the fitted peak value (also indicated by the vertical dashed line), FWHM (also indicated by the dotted
vertical lines), and signal-to-noise ratio.}
\label{fig:finalResults}
\end{figure*}

To prepare the x-derivative images for cross-correlation, their eastern regions, extending between Right Ascensions (J2000) of
17$^{\mbox{\scriptsize{h}}}$47$^{\mbox{\scriptsize{m}}}$15\farcsec8 and
17$^{\mbox{\scriptsize{h}}}$47$^{\mbox{\scriptsize{m}}}$16\farcsec5 (see Figure \ref{fig:allImages}), were extracted and
padded along each side with 50 pixels (2500~mas) of value zero. These cropped and padded x-derivative images for the 1993 and 2005
epochs were then cross-correlated with each other over a range of non-integer pixel shifts between $-$2500 and $+$2500~mas in both RA
and Dec. The resultant 2005$-$1993 cross-correlation map, which indicates the shift required to make the 1993 epoch colocate with the
2005 epoch, is shown in the left of Figure \ref{fig:finalResults}.

Note that the map in Figure \ref{fig:finalResults} incorporates trial shifts large enough to probe regions
where the cross-correlation falls to zero (corresponding to cross-correlation between signal and a source-free region,
as opposed to only probing trial shifts close to the maxima of each x-derivative map). In this way, the contours presented
in Figure \ref{fig:finalResults} represent percentages of the peak cross-correlation value.

To quantify the shift between epochs, a profile along a line parallel to the RA-axis was taken through the peak of the
cross-correlation map, as shown in the right of Figure \ref{fig:finalResults}. We assume that morphological changes
are negligible in the eastern region of the x-derivative images; therefore, by taking a profile through the peak we tolerate
small Dec shifts between the two epochs.

The RA shift between the 2005 and 1993 epochs was determined by fitting a Gaussian to the central 660~mas of the cross-correlation
profile. The resultant shift is 154~mas with a statistical uncertainty of 7~mas, where the latter is equal to the FWHM divided by
twice the signal-to-noise ratio. This calculation reflects the angular resolution and noise in the images, but to completely quantify
the error on the shift between the two epochs, systematic errors also need to be incorporated.

To estimate the positional error in the image plane, corresponding to phase error in the spatial frequency plane, the phases of the complex
visibility data for each epoch were self-calibrated\footnote{Note that self-calibration was not used in the general reduction process because
it would have caused a degradation in relative positional information between the final images, as the phases would no longer be tied to a
secondary calibrator of known position.}. By dividing the standard error of the mean of the phase variation in the gain solutions by
180\degree, the fraction of a diffraction-limited beam by which positions may have been in error in the image plane were calculated. By
multiplying this fraction with the diffraction limited beamwidth for the 1993 and 2005 epochs, the two-dimensional relative positional
uncertainty of these two reference frames was estimated to be 22 and 20 mas respectively.

The systematic error in our measurement, which describes the relative positional error between the two epochs, was then determined by
reducing the self-calibrated positional uncertainties for each epoch by $\sqrt{2}$ (we are only looking at random positional uncertainties
projected along the RA-axis), and added in quadrature. This error was found to be 21~mas, which totally dominates the much smaller
statistical error of 7~mas.

By calculating the total error as the quadrature sum of the statistical and systematic errors, the RA-shift of the PWN tip between the
1993 and 2005 epochs was found to be $154\pm22$~mas. When divided by the 4372 days elapsed between these two epochs, the measured shift
corresponds to a proper motion of $\mu=12.9\pm1.8$~mas~yr$^{-1}$ in an eastward direction. We therefore detect motion at the
$7\sigma$ level. Note that, at an assumed distance of $\sim$5~kpc along a line of sight toward the Galactic center, this motion is
unlikely to be contaminated significantly by Galactic rotation \citep{olling}.

If we simplistically compare the eastward component of the proper motion with the angle of the Mouse's cometary tail, as described earlier in
\S~\ref{SectionAnalysis}, we obtain a crude estimate of $\sim$1~mas~yr$^{-1}$ for the northerly component of the nebula's proper motion. As this
value is well within the error for the eastward motion, which is dominated by systematic effects, we feel that our earlier assumption of pure
eastward motion in the presence of relative positional uncertainty between the 1993 and 2005 reference frames is justified.

\section{Discussion}\label{SectionDiscussion}

Bow shock PWNe have a double-shock structure consisting of an outer bow shock where the ambient ISM is collisionally excited, an inner termination
shock at which the pulsar's relativistic wind is decelerated, and a contact discontinuity between these two shocks which marks the boundary between
shocked ISM and shocked pulsar wind material \citep{gaensler:3}. The outer bow shock may emit in H$\alpha$, though for systems such as
PWN~G359.23$-$0.82 with high levels of extinction along their line of sight, the detection of such emission would not be expected. The inner termination
shock, which encloses the pulsar's relativistic wind, may emit synchrotron radiation detectable at radio/X-ray wavelengths. It is expected that any
synchrotron emission beyond the termination shock would be sharply bounded by the contact discontinuity \citep{gaensler:5}.

As mentioned in \S~\ref{SectionAnalysis}, we suggest that the eastern regions of the x-derivative images from Figure \ref{fig:allImages}
provide the best opportunity to track motion of PSR~J1747$-$2958, relatively independent of any morphological changes occurring in
PWN~G359.23$-$0.82. Physically, these regions of greatest spatial derivative (along the RA-axis) might correspond to the vicinity of the
termination shock apex, or possibly the contact discontinuity between the two forward shocks, where motion of the pulsar is causing
confinement of its wind and where rapid changes in flux might be expected to occur over relatively small angular scales. This is
consistent with hydrodynamic simulations which predict that the apex of the bow shock will be located just outside a region of intense
synchrotron emission in which the pulsar lies \citep{bucc,van:1}.

The assumption that the eastern region of each x-derivative image can be used as a proxy to track the motion of PSR~J1747$-$2958 is
therefore plausible, but difficult to completely justify. To show that motion calculated in this way provides an upper limit to the
true motion of PSR~J1747$-$2958 we recall the overall morphological change described at the end of \S~\ref{SectionObservations}, namely
that the Mouse nebula has expanded with time between the 1993 to 2005 epochs. This expansion suggests that the ISM density may be
dropping, causing the termination shock to move further away from pulsar, so that any motion calculated using the nebula may in
fact overestimate the motion of the pulsar (a similar argument was used by \citeauthor{blazek} \citeyear{blazek} in placing an upper
limit on the motion of the PWN associated with PSR~B1757$-$24). Such changes in density are to be expected as the nebula moves through
interstellar space, where like the spectacular Guitar nebula \citep{chatterjee:1} motion may reveal small-scale inhomogeneities in the
density of the ISM. We therefore assume that our measurement of proper motion from \S~\ref{Subsection:FinalCalc} corresponds to an
upper limit on the true proper motion of PSR~J1747$-$2958.

\subsection{Space Velocity and Environment of PSR~J1747$-$2958}\label{Subsec:SpaceVel}

Using our proper motion result from \S~\ref{Subsection:FinalCalc} and the arguments for interpreting this motion as an upper limit
from \S~\ref{SectionDiscussion}, the projected eastward velocity of PSR~J1747$-$2958 is inferred to be
$V_{\mbox{\tiny{PSR,$\perp$}}}$~{\footnotesize $\leq$}~$\left(306\pm43\right)d_{5}$~km~s$^{-1}$. Given that no motion along the line
of sight or in Dec could be measured, we will assume that our estimate of $V_{\mbox{\tiny{PSR,$\perp$}}}$ approximates
the 3-dimensional space velocity $V_{\mbox{\tiny{PSR}}}$.

In a bow shock PWN, the pulsar's relativistic wind will be confined and balanced by ram pressure. Using our proper motion upper limit
(assuming $V_{\mbox{\tiny{PSR}}}$~$\approx$~$V_{\mbox{\tiny{PSR,$\perp$}}}$), the pressure balance relationship\footnote{This
relationship assumes a uniform density $\rho$ with typical cosmic abundances, expressed as $\rho=1.37n_{0}m_{H}$, where $m_{H}$ is the
mass of a hydrogen atom and $n_{0}$ is the number density of the ambient ISM.}
$V_{\mbox{\tiny{PSR}}}=305n_{0}^{-1/2}d_{5}^{-1}$~km~s$^{-1}$ from \S~4.4 of \citet{gaensler:5}, and Monte Carlo simulation, we find an
in situ hydrogen number density $n_{0}$~$\approx$~$\left(1.0_{-0.2}^{+0.4}\right)d_{5}^{-4}$~cm$^{-3}$ at 68\%
confidence, or $n_{0,95\%}$~$\approx$~$\left(1.0_{-0.4}^{+1.1}\right)d_{5}^{-4}$~cm$^{-3}$ at 95\% confidence. Our calculated density
$n_{0}$ implies a local sound speed of $\sim$5~km~s$^{-1}$, corresponding to motion through the warm phase of the ISM.

Our space velocity for PSR~J1747$-$2958 is comparable with other pulsars that have observed bow shocks \citep{chatterjee:2}, and is
consistent with the overall projected velocity distribution of the young pulsar population \citep{hobbs, faucher}.
We note that \citet{gaensler:5} estimated a proper motion and space velocity of $\approx$25~mas~yr$^{-1}$ and $\approx$600~km~s$^{-1}$,
respectively, which are a factor of two larger than the values determined in this paper. However, by halving their assumed sound speed
of 10~km~s$^{-1}$, their estimates of motion correspondingly halve.

We now use our proper motion and hydrogen number density results to resolve the question of association between PSR~J1747$-$2958
and SNR~G359.1$-$0.5, and to investigate the age and possible future evolution of this pulsar.

\subsection{Association with SNR~G359.1$-$0.5?}\label{Subsection:Association}

If PSR~J1747$-$2958 and the adjacent SNR~G359.1$-$0.5 are associated and have a common progenitor, then an age estimate for the system that is
independent of both distance and inclination effects is simply the time taken for the pulsar to traverse the eastward angular separation
between the explosion site inferred from the SNR morphology and its current location, at its inferred eastward proper motion
{\footnotesize $\lesssim$}~$\mu$. Assuming pulsar birth at the center of the SNR, the eastward angular separation between the center of
SNR~G359.1$-$0.5 from \citet{uchida} and the location of PSR~J1747$-$2958 from the timing solution by \citet{gaensler:5} is
found to be $\theta \sim 23'$, which would imply a system age of $\theta / \mu$~{\footnotesize $\gtrsim$}~110~kyr.

Given such a large age, and the unremarkable interstellar hydrogen number density at the (currently assumed) nearby Mouse
(from \S~\ref{Subsec:SpaceVel}), it would be difficult to argue why SNR~G359.1$-$0.5 has not dissipated and faded from view. Instead,
SNR~G359.1$-$0.5 appears to be a middle aged remnant $\sim$18~kyr old which continues to emit thermal X-rays \citep{bamba}. We conclude,
independent of distance estimates to either the pulsar or the SNR, that PSR~J1747$-$2958 is moving too slowly to be physically associated
with the relatively young SNR~G359.1$-$0.5.

\subsection{Age Estimate for PSR~J1747$-$2958}\label{Subsection:Age}

Given that an association between the Mouse and SNR~G359.1$-$0.5 is unlikely, as outlined in \S~\ref{Subsection:Association}, we now estimate the
age of PSR~J1747$-$2958 assuming that it is unrelated to this SNR.

As seen in Figure \ref{fig:yusef20cm}, there is a cometary tail of emission extending around 12$^{\prime}$ ($\sim$17$d_{5}$~pc) westward of
the Mouse, containing shocked pulsar wind material flowing back from the termination shock about PSR~J1747$-$2958 \citep{gaensler:5}. We begin
by simplistically assuming that this pulsar was born at the tail's western tip. By dividing the tail length by the proper motion
$\mu$, we estimate an age of $t_{\mbox{\tiny{tail}}}$~$\approx$~$56$~kyr. Note that this age is independent of any distance estimates to the
Mouse or of inclination effects. However, given that the tail appears to simply fade away rather than terminate suddenly, it is possible that tail
could be much longer, and thus that the system could be much older (by considering the upper limit arguments from \S~\ref{SectionDiscussion},
this system age may be even greater still). As discussed in \S~4.7 of \citet{gaensler:5}, it is unlikely that the Mouse or
its entire tail could still be located inside an unseen associated SNR, given that the tail is smooth and uninterrupted. In addition, the lack
of a rejuvenated SNR shell anywhere along the length of the tail (or indeed beyond it), such as that seen, for example, in the interaction
between PSR~B1951$+$32 and SNR~CTB80 \citep{van:2}, supports the conclusion that the Mouse's tail is located entirely in the ISM. Therefore,
the rim of the Mouse's unseen associated SNR must be located at a minimum angular separation of $\sim$12$^{\prime}$ west of the Mouse's current
location, implying that $t_{\mbox{\tiny{tail}}}$ is a lower limit on the time elapsed since PSR~J1747$-$2958 was in the vicinity of this rim.

To estimate the total age of PSR~J1747$-$2958 we thus need to incorporate the time taken for this pulsar to escape its associated
SNR and reach its current location, taking into account the continued expansion of the SNR following pulsar escape (which will sweep
up, and therefore shorten, part of the Mouse's tail initially located in the ISM). Using Monte Carlo simulation and following \citet{van:1},
we find that the time taken for a pulsar, traveling with velocity
$V_{\mbox{\tiny{PSR,$\perp$}}}$~{\footnotesize $\leq$}~$\left(306\pm43\right)d_{5}$~km~s$^{-1}$ through a typical interstellar environment
with constant hydrogen number density $n_{0}$ from \S~\ref{Subsec:SpaceVel}, to escape from its SNR (while
in the pressure-drive snowplow phase) of typical explosion energy $\sim$10$^{51}$~ergs, and leave behind a 12$^{\prime}$ tail in the ISM
which remains ahead of the expanding SNR, is $163_{-20}^{+28}$~kyr at 68\% confidence, or $163_{-39}^{+60}$~kyr at 95\% confidence. Note that
the errors quoted for this total time incorporate the error in $\mu$ and are only weakly dependent on
the uncertainty in distance $d_{5}$ (for comparison, when the distance to PSR~J1747$-$2958 is fixed at 5~kpc, the 68\% and 95\% confidence
intervals are reduced to $164_{-17}^{+22}$~kyr and $164_{-31}^{+52}$~kyr, respectively).

Assuming that PSR~J1747$-$2958 was created in such a supernova explosion, and noting that the pulsars's travel time in the ISM
$t_{\mbox{\tiny{tail}}}$ is a lower limit (even without taking into account the upper limit associated with $\mu$, as described earlier in
this section), we can thus establish a lower limit on the age of the pulsar of
$t_{\mbox{\tiny{total}}}$~{\footnotesize $\geq$}~$163_{-20}^{+28}$~kyr (68\% confidence). This lower limit is greater than 6 times the
characteristic age $\tau_{c}=$~25.5~kyr of PSR~J1747$-$2958, which was derived from its measured pulse $P$ and $\dot{P}$ \citep{camilo:1},
suggesting that, within the context of the characteristic approximation, the spindown properties of this pulsar deviate significantly from
magnetic dipole braking (see \S~\ref{Subsection:Evolution}).

Our result is similar to the age discrepancy previously claimed for PSR~B1757$-$24 \citep{gaensler:2}; however, ambiguity regarding 
association with SNR G5.4$-$1.2 presents difficulties with this claim \citep{thorsett,zeiger}. In comparison, given the relatively
simple assumptions made in this paper, PSR~J1747$-$2958 arguably provides the most robust evidence to date that some pulsars may be much
older than their characteristic age. We now discuss the potential implications of this age discrepancy with regard to the future evolution
of PSR~J1747$-$2958.

\subsection{Possible Future Evolution of PSR~J1747$-$2958}\label{Subsection:Evolution}

Pulsars are assumed to slow down in their rotation according to the spin-down relationship $\dot{\Omega}=-K\Omega^{n}$, where $\Omega$ is the
pulsar's angular rotation frequency, $\dot{\Omega}$ is the angular frequency derivative, $n$ is the braking index, and $K$ is a positive constant
that depends on the pulsar's moment of inertia and magnetic moment. By taking temporal derivatives of this spin-down relationship, the rate at
which the characteristic age $\tau_{c}$ changes with time can be expressed as $d\tau_{c}/dt$=$(n-1)/2$ (e.g., \citeauthor{lyne5}
\citeyear{lyne5}). Evaluating $d\tau_{c}/dt$ as the ratio between $\tau_{c}=$~25.5~kyr and the lower age limit $t_{\mbox{\tiny{total}}}$,
we estimate a braking index of $n$~{\footnotesize $\lesssim$}~$1.3$ for PSR~J1747$-$2958 (incorporating the error limits from
$t_{\mbox{\tiny{total}}}$ does not significantly affect this value). Given that magnetic dipole braking corresponds
to $n=3$ (the value assumed when calculating $\tau_{c}$), the smaller braking index calculated here indicates that either some form of
non-standard pulsar braking is taking place, or that standard magnetic dipole braking is occurring in the presence of evolution of the magnetic
field or of the moment of inertia (e.g., \citeauthor{blandford} \citeyear{blandford}; \citeauthor{camilo:0} \citeyear{camilo:0}).
If we adopt a constant moment of inertia and assume standard electromagnetic dipole braking, then by performing a similar
derivation from the spin-down relationship for the surface magnetic field, the magnetic field growth timescale can be expressed as
$B/\left(dB/dt\right)$=$\tau_{c}/\left(3-n\right)$ (e.g., \citeauthor{lyne5} \citeyear{lyne5}). Evaluating this with our braking index, the magnetic
field growth timescale\footnote{As noted by \citet{lyne5} the magnetic field growth is not linear over time, as can be appreciated by taking an
example of a pulsar with braking index $n$=1.} is estimated to be $\approx$15~kyr.

\begin{figure}[h]
\setlength{\abovecaptionskip}{-7pt}
\begin{center}
\includegraphics[trim = 0mm 0mm 0mm 0mm, clip, width=8.5cm, angle=-90]{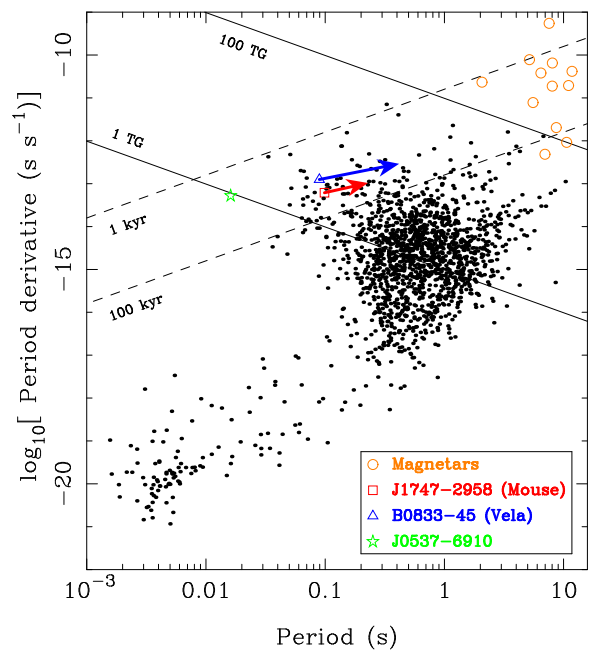}
\end{center}
\caption{The $P-\dot{P}$ diagram for rotating neutron stars, where points indicate the period and period derivative for over 1600 rotating
stars where such measurements were available (data obtained from the ATNF Pulsar Catalogue, version 1.34; \citeauthor{manchester} \citeyear{manchester}).
The periods and period derivatives of three neutron stars, B0833$-$45 (the Vela pulsar), J1747$-$2958 (the Mouse pulsar), and J0537$-$6910 (a 16-ms
pulsar in the Large Magellanic Cloud) are indicated with symbols (see discussion at the end of \S~\ref{Subsection:Evolution}), as well as those of
the known magnetars. Dashed lines indicate contours of constant $\tau_{c}$, while solid lines indicate contours of constant surface magnetic field
$B$. Points toward the bottom left are millisecond (recycled) pulsars, those in the central concentration are middle$-$old age pulsars, and those
to the top left are young pulsars. The middle$-$old age pulsars are presumed to move down to the right along lines of constant magnetic field,
corresponding to magnetic dipole braking, while some young pulsars such as the Vela pulsar have been found to be moving away from the main body of
pulsars, up and to the right toward the region of the magnetars. The plotted vectors for PSR~B0833$-$45 \citep{lyne5} and for PSR~J1747$-$2958 (this
paper) represent inferred trajectories across the diagram over 50~kyr, assuming a constant braking index over this time period; the magnitude of
each vector is proportional to the magnetic field growth timescale of the respective neutron star.}
\label{fig:ppdot}
\end{figure}

It is interesting to note that the braking index inferred here for PSR~J1747$-$2958 is comparable to the value obtained from an estimate
of $\ddot{P}$ made for the Vela pulsar B0833$-$45, which was found to have $n = 1.4 \pm 0.2$ \citep{lyne4}. To
investigate the possible future evolution of PSR~J1747$-$2958, we plot its implied trajectory (along with that of the Vela pulsar) across the
$P-\dot{P}$ diagram, as shown in Figure \ref{fig:ppdot}. Note that the magnitude of each plotted vector indicates motion
over a timescale of 50 kyr assuming that $\ddot{P}$ is constant, which is not true for constant $n$; however, the trend is apparent. The longer
vector for the Vela pulsar simply indicates that it is braking more rapidly than PSR~J1747$-$2958.

The plotted vectors for the Vela and Mouse pulsars both seem to point in the direction of the magnetars (high-energy neutron stars for which
$B${\footnotesize $\gtrsim$}10$^{14}$~G; for a review of magnetars, see \citeauthor{woods} \citeyear{woods}). By extrapolating the
trajectories in Figure \ref{fig:ppdot}, and assuming negligible magnetic field decay over time, it can be suggested that young, energetic,
rapidly spinning pulsars such as PSR~J0537$-$6910 \citep{marshall}, whose location in the $P-\dot{P}$ plane is shown with a star symbol,
may evolve into objects like the Vela or Mouse pulsars, which, as proposed by \citet{lyne5}, may in turn continue to undergo magnetic
field growth until arriving in the parameter space of the magnetars.

\section{Conclusions}\label{SectionConclusions}

We have investigated two epochs of interferometric data from the VLA spanning 12 years to indirectly infer a proper motion
for the radio pulsar J1747$-$2958 through observation of its bow shock PWN~G359.23$-$0.82. Derivative images were used to highlight
regions of rapid spatial variation in flux density within the original images, corresponding to the vicinity of the forward
termination shock, thereby acting as a proxy for the motion of the pulsar.

We measure an eastward proper motion for PWN~G359.23$-$0.82 of $\mu=12.9\pm1.8$~mas~yr$^{-1}$, and interpret this value as an upper
limit on the motion of PSR~J1747$-$2958. At this angular velocity, we argue that PSR~J1747$-$2958 is moving too slowly to be physically
associated with the relatively young adjacent SNR~G359.1$-$0.5, independent of distance estimates to either object or of inclination effects.

At a distance $d=5d_{5}$~kpc, the proper motion corresponds to a projected velocity of
$V_{\mbox{\tiny{PSR,$\perp$}}}$~{\footnotesize $\leq$}~$\left(306\pm43\right)d_{5}$~km~s$^{-1}$, which is consistent with the projected
velocity distribution for young pulsars. Combining the time taken for PSR~J1747$-$2958 to traverse its smooth $\sim$12$^\prime$ radio
tail with the time to escape a typical SNR, we calculate a lower age limit for PSR~J1747$-$2958 of
$t_{\mbox{\tiny{total}}}$~{\footnotesize $\geq$}~$163_{-20}^{+28}$~kyr (68\% confidence).

The lower age limit $t_{\mbox{\tiny{total}}}$ exceeds the characteristic age of PSR~J1747$-$2958 by more than a factor of 6, arguably providing
the most robust evidence to date that some pulsars may be much older than their characteristic age. This age discrepancy for PSR~J1747$-$2958
suggests that the pulsar's spin rate is slowing with an estimated braking index $n$~{\footnotesize $\lesssim$}~$1.3$ and that its
magnetic field is growing on a timescale $\approx$15~kyr. Such potential for magnetic field growth in PSR~J1747$-$2958,
in combination with other neutron stars that transcend their archetypal categories such as PSR~J1718$-$3718, a radio pulsar with a
magnetar-strength magnetic field that does not exhibit magnetar-like emission \citep{kaspi}, PSR~J1846$-$0258, a rotation-powered pulsar
that exhibits magnetar-like behaviour \citep{gavril,archibald}, and magnetars such as 1E 1547.0$-$5408 that exhibit radio
emission (\citeauthor{camilo:2} \citeyear{camilo:2}, and references therein), supports the notion that there may be evolutionary links
between the rotation-powered and magnetar classes of neutron stars. However, such a conclusion may be difficult to reconcile with evidence
suggesting that magnetars are derived from more massive progenitors than normal pulsars (e.g., \citeauthor{gaens:8} \citeyear{gaens:8};
\citeauthor{muno} \citeyear{muno}). If the massive progenitor hypothesis is correct, then this raises further questioning of whether, like
the magnetars, there is anything special about the progenitor properties of neutron stars such as PSR~J1747$-$2958, or whether all
rotation-powered pulsars exhibit similar magnetic field growth or even magnetar-like phases in their lifetimes.

To constrain the motion of PSR~J1747$-$2958 further, future observational epochs are desirable. It may be possible to better
constrain the motion and distance to this pulsar by interferometric astrometry with the next generation of sensitive
radio telescopes (e.g., \citeauthor{2004NewAR..48.1413C} \citeyear{2004NewAR..48.1413C}). High time resolution X-ray observations may
also be useful to detect any magnetar-like behaviour from this rotation-powered radio pulsar. In general, more neutron star
discoveries, as well as measured or inferred braking indices, may allow for a better understanding of possible neutron star evolution.

\acknowledgments

We thank the anonymous referee for their helpful comments.
C.~A.~H. acknowledges the support of an Australian Postgraduate Award and a CSIRO OCE Scholarship.
B.~M.~G. acknowledges the support of a Federation Fellowship from the Australian Research Council through grant FF0561298.
S.~C. acknowledges support from the University of Sydney Postdoctoral Fellowship program.
The National Radio Astronomy Observatory is a facility of the National Science Foundation
operated under cooperative agreement by Associated Universities, Inc.

{\it Facilities:} \facility{VLA}.


\begin{thebibliography}{}


\bibitem[Archibald et al.(2008)]{archibald} Archibald, A.~M., Kaspi, V.~M., Livingstone, M.~A., 
\& McLaughlin, M.~A.\ 2008, \apj, 688, 550 

\bibitem[Bamba et al.(2000)]{bamba} Bamba, A., Yokogawa, J., Sakano, M., \& Koyama, K.\ 2000, \pasj, 52, 259 

\bibitem[Blandford \& Romani(1988)]{blandford} Blandford, R.~D., \& Romani, R.~W.\ 1988, \mnras, 234, 57P

\bibitem[Blazek et al.(2006)]{blazek} Blazek, J.~A., Gaensler, B.~M., Chatterjee, S., van der Swaluw, E., Camilo, F., 
\& Stappers, B.~W.\ 2006, \apj, 652, 1523 

\bibitem[Braun et al.(1989)]{braun} Braun, R., Goss, W.~M., \& Lyne, A.~G.\ 1989, \apj, 340, 355 

\bibitem[Bucciantini(2002)]{bucc} Bucciantini, N.\ 2002, \aap, 387, 1066 

\bibitem[Camilo(1996)]{camilo:0} Camilo, F.\ 1996, in Pulsars: Problems and Progress, ed.
S. Johnston, M.~A. Walker, \& M. Bailes (San Francisco: ASP), 105, 39

\bibitem[Camilo et al.(2002)]{camilo:1} Camilo, F., Manchester, R.~N., Gaensler, B.~M., \& Lorimer, D.~R.\ 2002, \apjl, 579, L25 

\bibitem[Camilo et al.(2008)]{camilo:2} Camilo, F., Reynolds, J., Johnston, S., Halpern, J.~P., \& Ransom, S.~M.\ 2008,
\apj, 679, 681

\bibitem[Chatterjee \& Cordes(2002)]{chatterjee:2} Chatterjee, S., \& Cordes, J.~M.\ 2002, \apj, 575, 407 

\bibitem[Chatterjee \& Cordes(2004)]{chatterjee:1} Chatterjee, S., \& Cordes, J.~M.\ 2004, \apjl, 600, L51

\bibitem[Cordes et al.(2004)]{2004NewAR..48.1413C} Cordes, J.~M., Kramer, M., Lazio, T.~J.~W., Stappers, B.~W., Backer, D.~C., 
\& Johnston, S.\ 2004, New Astronomy Review, 48, 1413

\bibitem[Faucher-Gigu{\`e}re \& Kaspi(2006)]{faucher} Faucher-Gigu{\`e}re, C.-A., \& Kaspi, V.~M.\ 2006, \apj, 643, 332 

\bibitem[Gaensler et al.(2004)]{gaensler:5} Gaensler, B.~M., van der Swaluw, E., Camilo, F., Kaspi, V.~M., Baganoff, F.~K.,
Yusef-Zadeh, F., \& Manchester, R.~N.\ 2004, \apj, 616, 383 

\bibitem[Gaensler et al.(2005)]{gaens:8} Gaensler, B.~M., McClure-Griffiths, N.~M., Oey, M.~S., Haverkorn, M., Dickey, J.~M., 
\& Green, A.~J.\ 2005, \apjl, 620, L95 

\bibitem[Gaensler \& Frail(2000)]{gaensler:2} Gaensler, B.~M., \& Frail, D.~A.\ 2000, \nat, 406, 158 

\bibitem[Gaensler \& Slane(2006)]{gaensler:3} Gaensler, B.~M., \& Slane, P.~O.\ 2006, \araa, 44, 17 

\bibitem[Gavriil et al.(2008)]{gavril} Gavriil, F.~P., Gonzalez, M.~E., Gotthelf, E.~V., Kaspi, V.~M.,
Livingstone, M.~A., \& Woods, P.~M.\ 2008, Science, 319, 1802 

\bibitem[Gooch(1996)]{gooch} Gooch, R.\ 1996, in Astronomical Data Analysis Software and Systems V,
ed. G.~H.~Jacoby, \& J.~Barnes (San Francisco: ASP), 101, 80

\bibitem[Hobbs et al.(2005)]{hobbs} Hobbs, G., Lorimer, D.~R., Lyne, A.~G., \& Kramer, M.\ 2005, \mnras, 360, 974 

\bibitem[Kaspi \& McLaughlin(2005)]{kaspi} Kaspi, V.~M., \& McLaughlin, M.~A.\ 2005, \apjl, 618, L41

\bibitem[Lyne(2004)]{lyne5} Lyne, A.~G.\ 2004, in Young Neutron Stars and Their Environments, ed. F. Camilo \&
B.~M. Gaensler (San Francisco: ASP), 218, 257

\bibitem[Lyne et al.(1996)]{lyne4} Lyne, A.~G., Pritchard, R.~S., Graham-Smith, F., \& Camilo, F.\ 1996,
\nat, 381, 497

\bibitem[Manchester et al.(2005)]{manchester} Manchester, R.~N., Hobbs, G.~B., Teoh, A., \& Hobbs, M.\ 2005, \aj,
129, 1993, http://www.atnf.csiro.au/research/pulsar/psrcat/

\bibitem[Marshall et al.(1998)]{marshall} Marshall, F.~E., Gotthelf, E.~V., Zhang, W., Middleditch, J., 
\& Wang, Q.~D.\ 1998, \apjl, 499, L179 

\bibitem[Mori et al.(2005)]{mori} Mori, H., Maeda, Y., Pavlov, G.~G., Sakano, M., \& Tsuboi, Y.\ 2005,
Advances in Space Research, 35, 1137 

\bibitem[Muno et al.(2006)]{muno} Muno, M.~P., et al.\ 2006, \apjl, 636, L41

\bibitem[Olling \& Merrifield(1998)]{olling} Olling, R.~P., \& Merrifield, M.~R.\ 1998, \mnras, 297, 943 

\bibitem[Predehl \& Kulkarni(1995)]{predehl} Predehl, P., \& Kulkarni, S.~R.\ 1995, \aap, 294, L29 

\bibitem[Sault \& Killeen(2006)]{sault:2} Sault, R.~J., \& Killeen, N.~E.~B.\ 2006, The Miriad User's Guide
(Sydney: ATNF), http://www.atnf.csiro.au/computing/software/miriad/

\bibitem[Sankrit \& Hester(1997)]{sankrit} Sankrit, R., \& Hester, J.~J.\ 1997, \apj, 491, 796 

\bibitem[Thorsett et al.(2002)]{thorsett} Thorsett, S.~E., Brisken, W.~F., \& Goss, W.~M.\ 2002, \apjl, 573, L111 


\bibitem[Uchida et al.(1992)]{uchida} Uchida, K., Morris, M., \& Yusef-Zadeh, F.\ 1992, \aj, 104, 1533 

\bibitem[van der Swaluw et al.(2003)]{van:1} van der Swaluw, E., Achterberg, A., Gallant, Y.~A., Downes, T.~P.,
\& Keppens, R.\ 2003, \aap, 397, 913

\bibitem[van der Swaluw et al.(2002)]{van:2} van der Swaluw, E., Achterberg, A., \& Gallant, Y.~A.\ 2002, in Neutron
Stars in Supernova Remnants, ed. P.~O. Slane \& B.~M. Gaensler (San Francisco: ASP), 271, 135

\bibitem[Woods \& Thompson(2006)]{woods} Woods, P.~M., \& Thompson, C.\ 2006, in Compact stellar X-ray sources, ed. W.
Lewin \& M. van der Klis (Cambridge, UK: Cambridge University Press), 547

\bibitem[Yusef-Zadeh \& Bally(1987)]{yusef} Yusef-Zadeh, F., \& Bally, J.\ 1987, \nat, 330, 455 

\bibitem[Yusef-Zadeh \& Gaensler(2005)]{yusef3} Yusef-Zadeh, F., \& Gaensler, B.~M.\ 2005, Advances in Space Research, 35, 1129 

\bibitem[Zeiger et al.(2008)]{zeiger} Zeiger, B.~R., Brisken, W.~F., Chatterjee, S., \& Goss, W.~M.\ 2008, \apj, 674, 271

\end{thebibliography}
\end{document}